\documentclass[aps,prl,twocolumn,superscriptaddress,showpacs]{revtex4}

\usepackage{graphicx}
\usepackage{color}

\begin{document}
\title{Subband structure of a two-dimensional electron gas formed at the polar surface of the strong spin-orbit perovskite KTaO$_3$}

\author{P.~D.~C.~King}
\affiliation{SUPA, School of Physics and Astronomy, University of St. Andrews, St. Andrews, Fife KY16 9SS, United Kingdom}

\author{R.~H.~He}
\affiliation{Advanced Light Source, Lawrence Berkeley National Lab, Berkeley, CA 94720, USA}

\author{T.~Eknapakul}
\author{P. Buaphet}
\affiliation{School of Physics, Suranaree University of Technology and Synchrotron Light Research Institute, Nakhon Ratchasima, 30000, Thailand}

\author{S.-K.~Mo}
\affiliation{Advanced Light Source, Lawrence Berkeley National Lab, Berkeley, CA 94720, USA}

\author{Y.~Kaneko}
\affiliation{Multiferroics Project, ERATO, JST, Tokyo 113-8656, Japan}

\author{S.~Harashima}
\affiliation{Department of Applied Physics, University of Tokyo, Bunkyo-ku, Tokyo 113-8656, Japan}

\author{Y.~Hikita}
\affiliation{Departments of Physics and Applied Physics, Stanford University, CA 94305, USA} 
\affiliation {SIMES, SLAC National Accelerator Laboratory, 2575 Sand Hill Road, CA 94025, USA}

\author{M.~S.~Bahramy}
\affiliation{Correlated Electron Research Group (CERG), RIKEN-ASI, Wako 351-0918, Japan}

\author{C.~Bell}
\affiliation{Departments of Physics and Applied Physics, Stanford University, CA 94305, USA} 
\affiliation {SIMES, SLAC National Accelerator Laboratory, 2575 Sand Hill Road, CA 94025, USA}

\author{Z.~Hussain}
\affiliation{Advanced Light Source, Lawrence Berkeley National Lab, Berkeley, CA 94720, USA}

\author{Y.~Tokura}
\affiliation{Multiferroics Project, ERATO, JST, Tokyo 113-8656, Japan}
\affiliation {Department of Applied Physics, University of Tokyo, Bunkyo-ku, Tokyo 113-8656, Japan}
\affiliation{Correlated Electron Research Group (CERG), RIKEN-ASI, Wako 351-0918, Japan}

\author{Z.-X.~Shen}
\affiliation{Departments of Physics and Applied Physics, Stanford University, CA 94305, USA} 
\affiliation {SIMES, SLAC National Accelerator Laboratory, 2575 Sand Hill Road, CA 94025, USA}

\author{H.~Y.~Hwang}
\affiliation{Departments of Physics and Applied Physics, Stanford University, CA 94305, USA} 
\affiliation {SIMES, SLAC National Accelerator Laboratory, 2575 Sand Hill Road, CA 94025, USA}
\affiliation{Correlated Electron Research Group (CERG), RIKEN-ASI, Wako 351-0918, Japan}

\author{F.~Baumberger}
\email[Corresponding e-mail: ]{fb40@st-andrews.ac.uk}
\affiliation {SUPA, School of Physics and Astronomy, University of St. Andrews, St. Andrews, Fife KY16 9SS, United Kingdom}

\author{W. Meevasana}
\email[Corresponding e-mail: ]{worawat@g.sut.ac.th}
\affiliation{School of Physics, Suranaree University of Technology
and Synchrotron Light Research Institute, Nakhon Ratchasima,
30000, Thailand} \affiliation{Thailand Center of Excellence in
Physics, CHE, Bangkok, 10400, Thailand}

\date{\today}
\begin{abstract}
We demonstrate the formation of a two-dimensional electron gas (2DEG) at the $(100)$ surface of the $5d$ transition-metal oxide KTaO$_3$. From angle-resolved photoemission, we find that quantum confinement lifts the orbital degeneracy of the bulk band structure and leads to a 2DEG composed of ladders of subband states of both light and heavy carriers. Despite the strong spin-orbit coupling, our measurements provide a direct upper bound for potential Rashba spin splitting of only $\Delta{k}_\parallel\sim\!0.02$~\AA$^{-1}$ at the Fermi level. The polar nature of the KTaO$_3(100)$ surface appears to help mediate formation of the 2DEG as compared to non-polar SrTiO$_3(100)$.
\end{abstract}

\maketitle

Today's electronic devices largely rely on the tuneability of narrow conducting channels in semiconductor hosts. Creating such two-dimensional electron gases (2DEGs) in oxides, which in bulk form generally show much larger and more diverse responses to external stimuli, holds the potential for devices with functionalities well beyond what we have experienced to date~\cite{Takagi:Science:327(2010)1601--1602,Mannhart:Science:327(2010)1607--1611}. The prototypical oxide 2DEG is formed when SrTiO$_3$ is interfaced to the polar surface of another perovskite oxide~\cite{Ohtomo:Nature:427(2004)423--426}. This system indeed shows novel properties~\cite{Mannhart:Science:327(2010)1607--1611,Zubko:Annu.Rev.Condens.MatterPhys.:2(2011)141--165}, such as an unusual co-existence of ferromagnetism and superconductivity~\cite{Bert:NaturePhys.:7(2011)767--771,Li:NaturePhys.:7(2011)762--766}. Several combinations of ABO$_3$/SrTiO$_3$ heterostructures have been investigated to date, incorporating both Mott~\cite{Hotta:Phys.Rev.Lett.:99(2007)236805} and band insulators~\cite{Ohtomo:Nature:427(2004)423--426,Kalabukhov:arXiv:0704.1050:(2007)} as the overlayer. However, the 2DEGs formed were always found to reside in SrTiO$_3$. There is great current interest in inducing 2DEGs in more exotic parent materials~\cite{Oshima}. The recent discovery that oxygen vacancies mediate formation of a 2DEG at the bare surface of SrTiO$_3$~\cite{Meevasana:NatureMater.:10(2011)114--118,Santander-Syro:Nature:469(2011)189--193} may provide a route to achieve this. 

Of particular interest are $5d$ transition metal oxides (TMOs). Their large spin-orbit interactions are thought to drive the formation of a host of unconventional ground states such as $J=1/2$ Mott insulators~\cite{Kim:Phys.Rev.Lett.:101(2008)076402,Kim:Science:323(2009)1329--1332}, correlated topological insulators~\cite{Shitade:Phys.Rev.Lett.:102(2009)256403,Pesin:NaturePhys.:6(2010)376--381}, and spin-triplet superconductors~\cite{Yi-Zhuang_You}. Moreover, $5d$ TMOs offer the potential to incorporate the spintronic functionality sought in emerging schemes of semiconductor electronics~\cite{Zutic:Rev.Mod.Phys.:76(2004)323--410,Koo:Science:325(2009)1515--1518,Nadj-Perge:Nature:468(2010)1084--1087} into all-oxide devices. They could therefore provide a novel and potentially very rich host for engineering of artificial 2D electron systems. Understanding the interplay of strong spin-orbit coupling, quantum confinement, and electronic correlations within such a 2DEG is an essential step towards realizing their potential for practical applications. To date, however, this has been hampered by the difficulty of generating 2DEGs localized in $5d$ oxides via interface engineering.

Here, we show that such a 2DEG can be created at the $(100)$ surface of the $5d$ perovskite KTaO$_3$. We utilize angle-resolved photoemission (ARPES) to provide the first direct measurement of the subband structure of a $5d$-oxide 2DEG. Our model calculations show quantitative agreement with ARPES measurements of both SrTiO$_3$ and KTaO$_3$ 2DEGs, strongly constraining theoretical pictures of such systems. In particular, we find a delicate interplay of multi-orbital physics, quantum confinement and spin-orbit interactions, driving orbital ordering within the 2DEG. Surprisingly, however, the 2DEG does not exhibit the large Rashba spin-splitting which might naturally be expected.

Single-crystal undoped KTaO$_3$ (commercial samples from Crystal Base Jpn.) and lightly electron-doped K$_{1-x}$Ba$_x$TaO$_3$ (flux-grown samples, $x < 0.001$) was measured. The Ba-doping yields a small residual bulk conductivity ($n\sim\!1\times10^{19}$~cm$^{-3}$ from Hall effect measurements) which eliminates charging effects in ARPES, but does not otherwise affect the conclusions of this work. ARPES measurements ($T=20$~K, $h\nu=45-85$~eV) were performed using a Scienta R4000 hemispherical analyzer at beamline 10.0.1 of the Advanced Light Source with an energy resolution between 8 and 35 meV, and an angular resolution of 0.35$^\circ$. Multiple samples were cleaved along the $(100)$ surface at the measurement temperature in a pressure better than $3\times10^{-11}$~mbar.

\begin{figure}
\begin{center}
\includegraphics[width=\columnwidth]{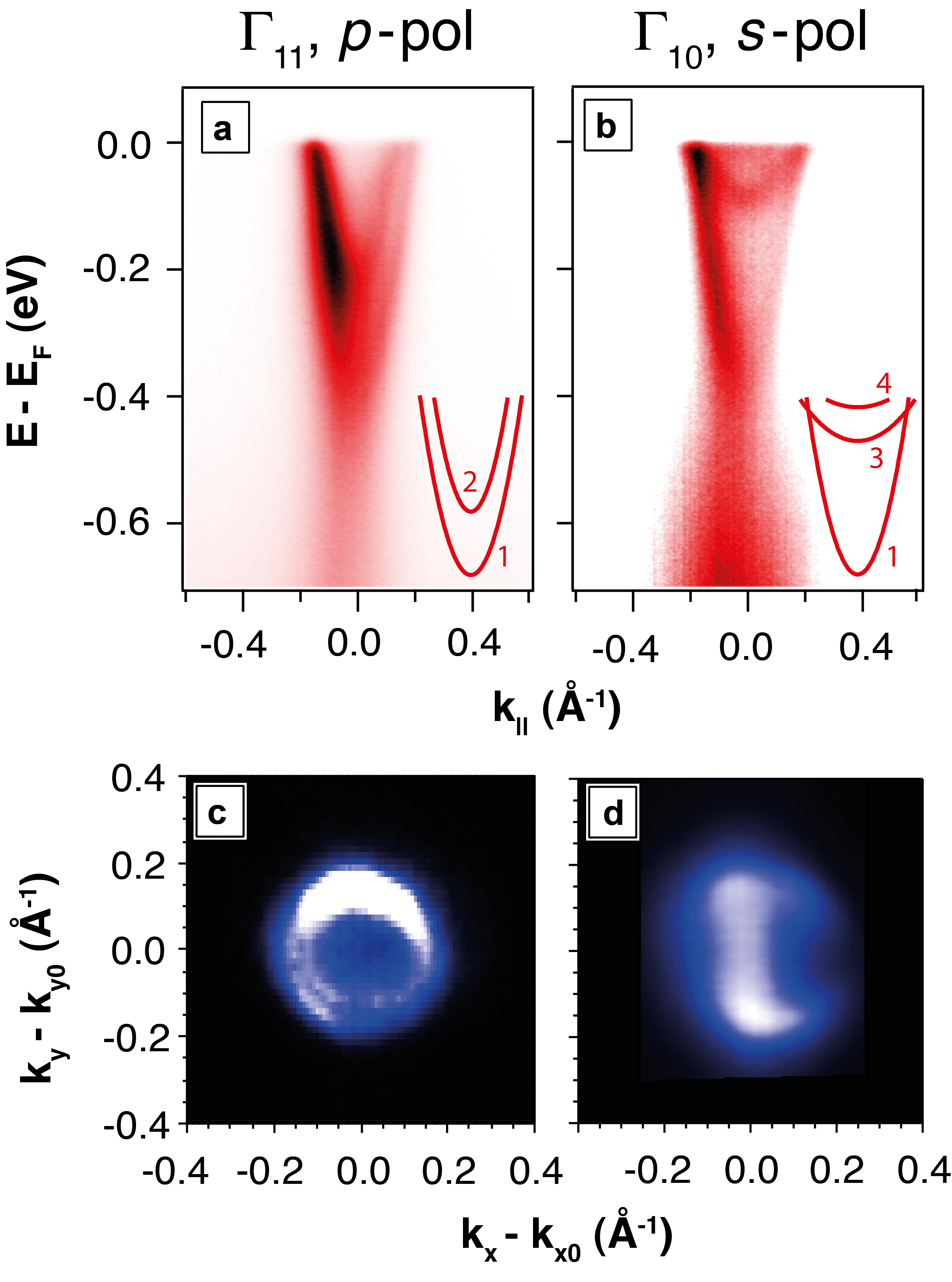}
\caption{ \label{f:2DEG} (a,b) ARPES measurements of the $\Gamma$--X dispersion of surface 2DEG states in KTaO$_3$, measured using $p$- and $s$-polarized $55$~eV synchrotron light around the $\Gamma_{11}$ and $\Gamma_{10}$ points, respectively. A schematic representation of the measured band structure is shown inset. (c,d) Equivalent measurements of the Fermi surface. }
\end{center}
\end{figure}

Our surface-sensitive ARPES measurements (Fig.~\ref{f:2DEG}) reveal a complex electronic structure with at least 4 dispersive electron-like bands which cross the chemical potential (Fig.~\ref{f:2DEG}(a,b)). This directly indicates that the surface of this material has become strongly conducting, in contrast to its bulk. We note that these measurements were performed at a photon energy chosen to probe electronic states near to the Brillouin zone boundary along $k_z$, where no bulk bands are expected in the vicinity of the Fermi level. Moreover, the same states are observed for both lightly bulk-doped and insulating undoped KTaO$_3$ samples (not shown). These states have equal Fermi wavevectors and occupied bandwidth (within our experimental resolution), even though the bulk carrier density should vary by at least 5 orders of magnitude between the samples. This conclusively rules out bulk states as the source of the measured bands. Photon-energy-dependent measurements (not shown) further reveal that the observed states have no dispersion along $k_z$. They are therefore two-dimensional electronic states confined near the surface, unlike the three-dimensional bulk states. 

\begin{figure}
\begin{center}
\includegraphics[width=\columnwidth]{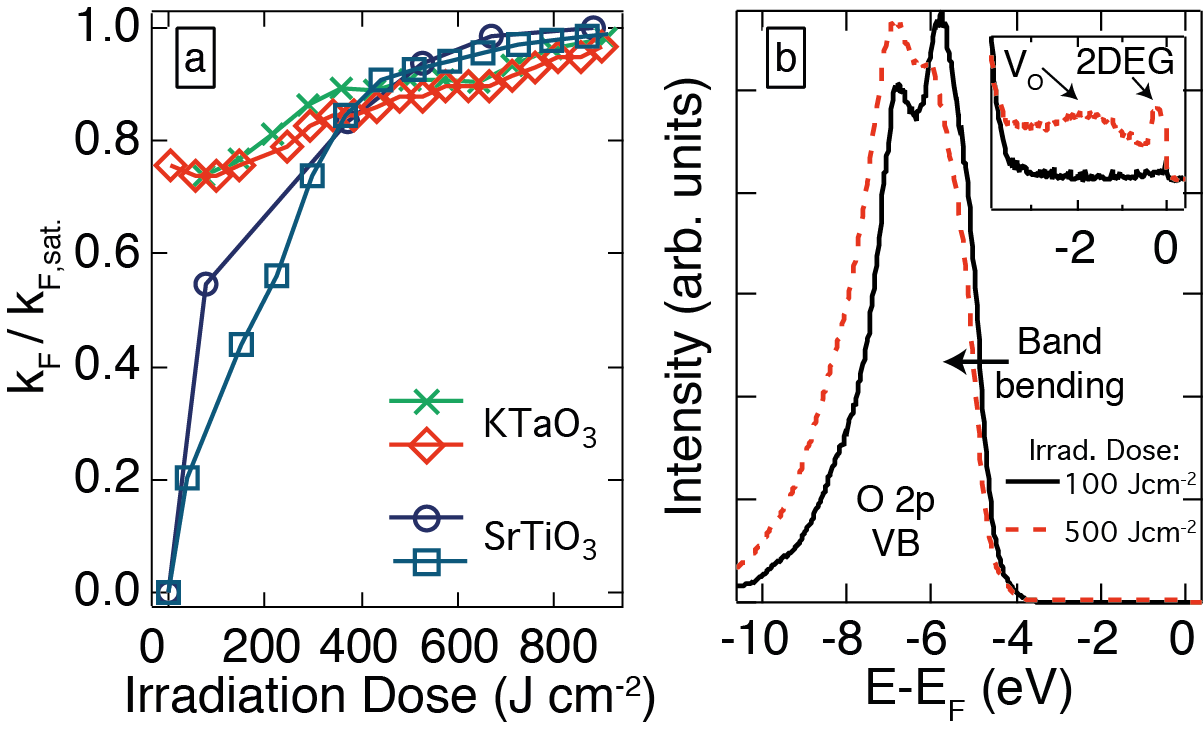}
\caption{ \label{f:VO} (a) Irradiation dose-dependence of Fermi wavevector of the deepest $d_{xy}$-type band of surface 2DEGs formed in SrTiO$_3$ and KTaO$_3$ upon exposure to intense UV light, normalized to their saturation values. (b) O~2$p$ valence bands of KTaO$_3$ after small and heavier irradiation dose, with the near-E$_F$ emission magnified in the inset. }
\end{center}
\end{figure}

After cleaving the sample, and upon exposure to intense UV light, the Fermi wavevectors of the states increase (Fig.~\ref{f:VO}(a)) and then saturate to give the electronic structure shown in Fig.~\ref{f:2DEG}. At the same time, the O~$2p$ valence bands shift to higher binding energy (Fig.~\ref{f:VO}(b)), indicating a downward bending of the valence and conduction bands relative to the Fermi level in the vicinity of the surface. This causes a build-up of charge near the surface~\cite{King:Phys.Rev.Lett.:101(2008)116808}. In conventional semiconductors, these electrons do not occupy the original bulk electronic states: the electrostatic band bending potential, together with the potential step at the surface itself, forms a quantum well. This causes the conduction bands to reconstruct into ladders of partially-filled two-dimensional subbands~\cite{King:Phys.Rev.Lett.:104(2010)256803}. The two-dimensional metallic states that we measure here by ARPES are the first direct observation of such quantum-confined states in a $5d$ TMO. The formation of the 2DEG is accompanied by the emergence of an in-gap defect peak (V$_\mathrm{O}$ in the inset of Fig.~\ref{f:VO}(b)), indicative of oxygen vacancies at the surface. This suggests that the density of the 2DEG in KTaO$_3$ may be tuned by controlling the positive surface charge resulting from a UV-stimulated desorption of oxygen, as recently found for a surface 2DEG created in SrTiO$_3$~\cite{Meevasana:NatureMater.:10(2011)114--118}.

However, unlike in SrTiO$_3$, the 2DEG in KTaO$_3$ exists right from the initial stages of our measurement (Fig.~\ref{f:VO}(a)). This is most likely due to the polar nature of KTaO$_3(100)$. In order to avoid the large energy cost associated with a polar catastrophe, KTaO$_3$ likely cleaves leaving approximately half a $($KO$)^-$ layer on the $($TaO$_2)^+$ surface. This structural arrangement may lower the formation energy for oxygen vacancies compared to SrTiO$_3(100)$, allowing much more rapid initial formation of the 2DEG. Alternately, if the surface TaO$_2$ plane is terminated by less than half a KO layer, the 2DEG could exist directly following the sample cleave in order to screen the intrinsic positive charge associated with the polar surface. Instead of the creation of oxygen vacancies, the increase in 2DEG density with irradiation dose could also be associated with desorption of $($KO$)^-$ from the surface. This would drive the system back towards an ideal polar surface. Intriguingly, from a comparison of model calculations (discussed below) to the ARPES measurements, we estimate the saturated density of the 2DEG to be $N\approx2\times10^{14}$~cm$^{-2}$, which is slightly lower than, but approaching, the 0.5$e^-$ per unit cell ($3.3\times10^{14}$~cm$^{-2}$) which would be expected from a simple polar catastrophe argument~\cite{Nakagawa:NatureMater.:5(2006)204--209} for a stoichiometric TaO$_2$ surface. Thus, while an interface between a polar and non-polar surface does not always appear necessary to create an oxide 2DEG~\cite{Meevasana:NatureMater.:10(2011)114--118,Santander-Syro:Nature:469(2011)189--193}, these measurements suggest that it may help mediate its formation, either via intrinsic electronic reconstruction~\cite{Nakagawa:NatureMater.:5(2006)204--209} or by promoting the formation of extrinsic defects.

The two highest binding energy bands of the resulting 2DEG (bands 1 and 2, Fig.~\ref{f:2DEG}(a)) have a light effective mass of $\sim\!0.3$~m$_e$, obtained from parabolic fits to their measured dispersion. This is almost a factor of 2 smaller than recently determined for a surface 2DEG in SrTiO$_3$~\cite{Meevasana:NatureMater.:10(2011)114--118}, suggesting that KTaO$_3$ could provide a superior platform to its workhorse counterpart of SrTiO$_3$ with which to develop high-mobility oxide electronics. We also observe a tail of intensity below the band bottom of these states, characteristic of the spectral function in strongly-interacting systems. This hints at an important role of electron correlations in this system, signifying a liquid-like state as was recently inferred for 2DEGs at SrTiO$_3$ surfaces and interfaces~\cite{Meevasana:NatureMater.:10(2011)114--118,Breitschaft:Phys.Rev.B:81(2010)153414}. Thus, KTaO$_3$-based 2DEGs may provide an appealing route to combine the exotic phase diagrams that often accompany strong electron correlations with a system which supports very mobile carriers, useful for device applications. In this respect, we note that a record mobility for a TMO 2DEG of 7000~cm$^2$V$^{-1}$s$^{-1}$ was recently achieved in a KTaO$_3$ electric double-layer transistor~\cite{Ueno:NatureNano.:advanceonlinepublication(2011)--}. The same system was also found to superconduct, a property which has not to date been obtained in the bulk of this material.

\begin{figure}
\begin{center}
\includegraphics[width=\columnwidth]{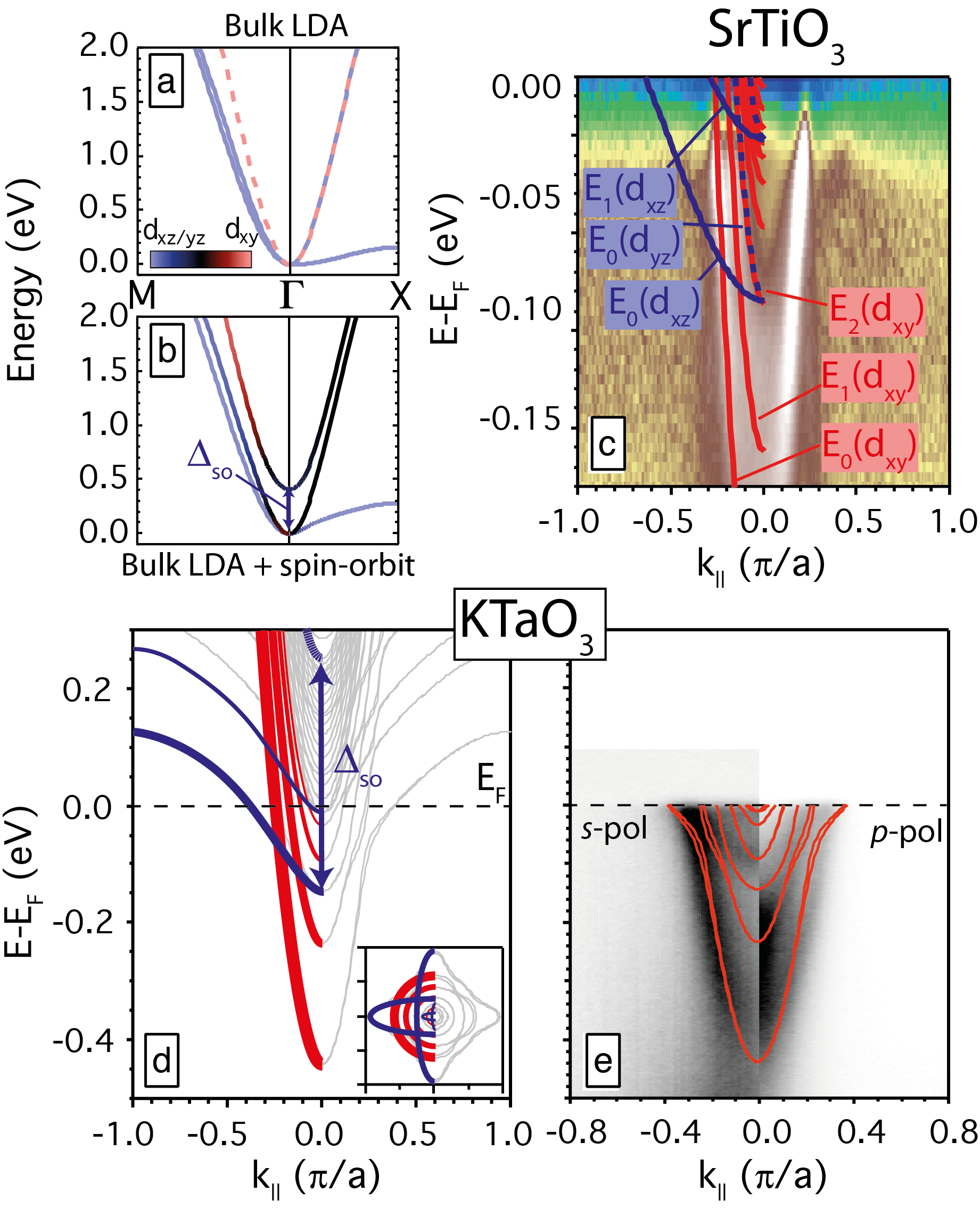}
\caption{ \label{f:model} LDA calculations of the bulk electronic structure and orbital character of KTaO$_3$ (a) excluding and (b) including spin-orbit coupling. (c) Comparison of measured dispersions to model tight-binding calculations of a surface 2DEG in SrTiO$_3$. (d) Equivalent calculations for a KTaO$_3$ 2DEG, including the strong spin-orbit coupling. The coloured lines give a schematic decomposition of its orbital makeup. (e) Comparison of the KTaO$_3$ calculations to the experimental data from Fig,~\ref{f:2DEG}(a,b).}
\end{center}
\end{figure}

Co-existing with these mobile states, we additionally observe much heavier carriers (bands 3 and 4 in Fig.~\ref{f:2DEG}(b), $m^*\sim\!2-3$~m$_e$). Together, the light and heavy mass bands contribute both concentric circular (Fig.~\ref{f:2DEG}(c)) as well as elliptical (Fig.~\ref{f:2DEG}(d)) electron pockets to the Fermi surface, suggesting that the electronic structure observed here is derived from multiple orbitals of different symmetries. Indeed, the bulk conduction bands of KTaO$_3$, as in SrTiO$_3$, are formed from three $t_{2g}$ orbitals~\cite{Neumann:Phys.Rev.B:46(1992)10623--10628}. In the simplest picture, these form three $d_{xy}$-, $d_{xz}$- and $d_{yz}$-derived interpenetrating ellipsoids, giving rise to one heavy and two light bands along the $\langle100\rangle$ directions (Fig.~\ref{f:model}(a)). However, spin-orbit coupling in KTaO$_3$ leads to a strong orbital hybridization. As shown in Fig.~\ref{f:model}(b), this lifts the $\Gamma$-point degeneracy by splitting off a light band above a pair of light and heavy bands by a large energy gap of $\Delta_{so}\sim\!400$~meV. The electronic bands we observe, however, have quite different characteristics (Fig.~\ref{f:2DEG}), with at least two light bands located at higher binding energies than the first heavy state. 

In the following, we show that this can be attributed to a modification of the orbital occupancy due to quantum confinement. We start by discussing the simpler case of a SrTiO$_3$ surface-2DEG (Fig.~\ref{f:model}(c)), where the spin-orbit split-off energy is small and can be neglected to first approximation. We model the electronic structure using a tight-binding supercell with band bending included via additional on-site potential terms, similar to the method introduced by Stengel~\cite{Stengel:Phys.Rev.Lett.:106(2011)136803}. This model is solved self-consistently with Poisson's equation, incorporating an electric-field-dependent dielectric constant~\cite{Copie:Phys.Rev.Lett.:102(2009)216804}, to yield the electronic structure shown in Fig.~\ref{f:model}(c).

Starting at the highest binding energies, a ladder of multiple $d_{xy}$-derived subbands are predicted. These are in good agreement with the multiple light states observed by ARPES (Fig.~\ref{f:model}(c)). This orbital assignment is consistent with their circular Fermi surfaces~\cite{Meevasana:NatureMater.:10(2011)114--118} and with the ladder of isotropic states recently observed in quantum oscillation measurements of $\delta$-doped SrTiO$_3$ quantum wells~\cite{Kozuka:Nature:462(2009)487--490,Kim:arXiv:1104.3388:(2011)}. Due to the small interlayer hopping between $d_{xy}$ orbitals along the confinement direction, or equivalently their heavy effective mass along $k_z$, the most deeply bound of these subbands have wavefunctions which are dominantly localized on successive atomic planes below the surface, explaining why they can be clearly observed in ARPES~\cite{note1}.  

On the other hand, the $d_{xz/yz}$ orbitals have a significantly larger overlap along $k_z$, leading to a much lighter effective mass along this direction. The binding energy of the heavy and light pairs of $d_{xz/yz}$-derived subbands is correspondingly reduced~\cite{Santander-Syro:Nature:469(2011)189--193}, lifting the orbital degeneracy of the bulk band structure~\cite{Salluzzo:Phys.Rev.Lett.:102(2009)166804}. This effect can be seen in our measurements, where a heavy band, whose binding energy and dispersion are in good agreement with the calculated $d_{xz}$-derived subband, can just be resolved in the normalized spectrum shown in Fig.~\ref{f:model}(c). Their reduced confinement energy results in subbands with envelope wavefunctions which are much more extended along $k_z$ than for the lower $d_{xy}$ bands, explaining the very weak spectral weight of the heavy bands in our measurements. The quantitative agreement of these calculations with the ARPES data confirms that the 2DEG in SrTiO$_3$ is generated by a near-surface band bending. This strongly supports first-principles calculations for the LaAlO$_3$/SrTiO$_3$ interface system that similarly find multiple co-existing confined and extended states within the 2DEG~\cite{Popovi'c:Phys.Rev.Lett.:101(2008)256801}, and constrains theoretical models of orbital ordering and degeneracy in these systems.
 
Our calculations can be readily adapted to KTaO$_3$. To the best of our knowledge, no prior calculations have considered the effects of a strong spin-orbit coupling on an oxide 2DEG. We show such calculations in Fig.~\ref{f:model}(d)~\cite{note2}. These can be qualitatively understood starting from the same orbital makeup as in SrTiO$_3$, but with some important additional features: (1) the lighter of the original $d_{xz/yz}$-derived states is shifted above the Fermi level by the large spin-orbit split-off energy, $\Delta_{so}$. This lifts the $\Gamma$-point degeneracy of $d_{xz/yz}$ states that is present for SrTiO$_3$; (2) small hybridization gaps open up between the different subbands; and (3) the orbital character of the lighter bands becomes strongly mixed. These characteristics are fully consistent with the measured band structure shown in Fig.~\ref{f:model}(e), although some of the more subtle features of the calculations cannot easily be resolved experimentally.

The calculations also predict a small spin splitting of the 2DEG states around the hybridization gaps (Fig.~\ref{f:model}(d,e)). This can be attributed to the Rashba effect, which lifts spin degeneracy in the presence of a structural inversion asymmetry~\cite{Bychkov:JETPLett.:39(1984)78}. The symmetry breaking is provided here by the asymmetric potential well which confines the 2DEG. However, despite the strong spin-orbit interactions, the calculated spin splitting $\Delta{k}_\parallel$ is only $\sim\!0.01$~\AA$^{-1}$ at the Fermi level. This is almost an order of magnitude smaller than recently observed for the seemingly similar system of a 2DEG in the heavy Bi-containing topological insulator Bi$_2$Se$_3$~\cite{King:arXiv:1103.3220:(2011)}, despite the much larger near-surface potential gradient confining the KTaO$_3$ 2DEG. We note, however, that a small Rashba splitting of 0.01~\AA$^{-1}$ is consistent with spin precession lengths extracted from weak antilocalization measurements of a KTaO$_3$ field-effect transistor~\cite{Nakamura:Phys.Rev.B:80(2009)121308}. This is also consistent with our experimental data, where any Rashba splitting is too small to be resolved, placing a direct, model-independent, upper bound of $\sim\!0.02$~\AA$^{-1}$ for spin-splitting at the Fermi level.

We attribute the small magnitude of the spin splitting to the particular electronic states involved: the degenerate $t_{2g}$ manifold of states at $\Gamma$ is split into an effective $J=3/2$ doublet at the conduction band edge and a $J=1/2$ split-off band (Fig.~\ref{f:model}(b))~\cite{Kim:Phys.Rev.Lett.:101(2008)076402}. This is analogous to the valence, rather than conduction, bands of typical III-V semiconductors such as GaAs. In such systems, the $k$-linear term in the Rashba spin-splitting of a 2D hole gas is forbidden due to symmetry, leaving the leading-order term as $k^3$~\cite{Winkler::(2003)}. In the low-$k$ regime applicable here, this yields a very small Rashba splitting even with the strong spin-orbit coupling and  large potential gradient within the 2DEG. 

\

This work was supported by the UK EPSRC (EP/F006640/1), ERC (207901), Scottish Funding Council, SUT Research and Development Fund, Suranaree University of Technology, the Japan Society for the Promotion of Science (JSPS) through its Funding Program for World-Leading Innovative R\&D on Science and Technology (FIRST Program), and the US Department of Energy, Office of Basic Energy Sciences, under contracts DE-AC02-76SF00515, DE-AC02-05CH11231, and DE-AC03-76SF00098. W.M. thanks M.F. Smith and S. Limpijumnong for helpful discussions.


\begin{thebibliography}{28}
\bibitem{note1}
  As the potential becomes shallower, the wavefunctions of the highest $d_{xy}$ states within this ladder become much more delocalized along $k_z$, making it difficult to resolve these predicted states in the ARPES measurements.
\end{thebibliography}
\end{document}